\documentclass[showpacs,amsmath,amssymb,aps,twocolumn,prl]{revtex4-1}%,superscriptaddress
\usepackage{graphicx}
\usepackage{soul}
\usepackage[colorlinks=true,citecolor=blue,linkcolor=magenta]{hyperref}
\usepackage[usenames]{color}
\usepackage[cspex,bbgreekl]{mathbbol}
\usepackage{relsize}

\newcommand {\grsim} {\ {\raise-.5ex\hbox{$\buildrel>\over\sim$}}\ }
\newcommand {\lessim} {\ {\raise-.5ex\hbox{$\buildrel<\over\sim$}}\ }

\begin{document}

\title{Relaxation dynamics in the merging of \textit{N} independent condensates}

\author{M.~Aidelsburger, J.~L.~Ville, R.~Saint-Jalm, S.~Nascimb\`ene, J.~Dalibard, J.~Beugnon}

\affiliation{Laboratoire Kastler Brossel, Coll\`ege de France, CNRS, ENS-PSL Research University, UPMC-Sorbonne Universit\'es, 11 place Marcelin-Berthelot, 75005 Paris, France}

\begin{abstract} 
Controlled quantum systems such as ultracold atoms can provide powerful platforms to study non-equilibrium dynamics of closed many-body quantum systems, especially since a complete theoretical description is generally challenging. In this Letter, we present a detailed study of the rich out-of-equilibrium dynamics of an adjustable number $N$ of uncorrelated condensates after connecting them in a ring-shaped optical trap. We observe the formation of long-lived supercurrents and confirm the scaling of their winding number with $N$ in agreement with the geodesic rule. Moreover, we provide insight into the microscopic mechanism that underlies the smoothening of the phase profile.
\end{abstract}

\maketitle

Thermalization of closed out-of-equilibrium many-body systems lies at the heart of statistical physics. Due to the recent progress in the preparation of well-controlled isolated quantum systems, this question can now be revisited in a quantum context \cite{Eisert:2015ka}. Whereas most systems are expected to reach thermal equilibrium, non-trivial situations can occur in integrable systems \cite{Langen:2016bu}, in the presence of disorder \cite{Nandkishore:2015kt} or due to the formation of long-lived topological defects \cite{Kibble:2007eq,DelCampo:2014eu}. Out-of-equilibrium dynamics are also central to the study of dynamical crossings of phase transitions. Indeed, the divergence of the relaxation time at the critical point for a second-order phase transition entails that the system cannot follow adiabatically the external perturbation. The relaxation dynamics can be used in that case to determine the critical exponents of the phase transition \cite{DelCampo:2014eu}. 

A rich situation occurs when $N$ condensates, characterized by independent initial phase factors, are coupled together.
Let us consider, for instance, the case where the condensates are placed along a ring and connections are suddenly established between neighboring condensates. One expects that, after some transient dynamics, stochastic metastable supercurrents are formed. This ring geometry was put forward by Zurek in a seminal paper \cite{Zurek:1985ko} drawing a parallel between laboratory experiments with liquid helium and classes of early universe theories. More recently this gedankenexperiment inspired experiments with superconducting loops \cite{Carmi:2000bv,Monaco:2002bd,Monaco:2009hg} and cold atoms \cite{Corman:2014cm}. A key ingredient of Zurek's study is the relation between the winding number of the supercurrent and the number of initial condensates $N$ according to the \textit{geodesic rule}. In essence, it enables a computation of the winding number based on the minimization of the kinetic energy of the system.  

\begin{figure}[t!]
\includegraphics{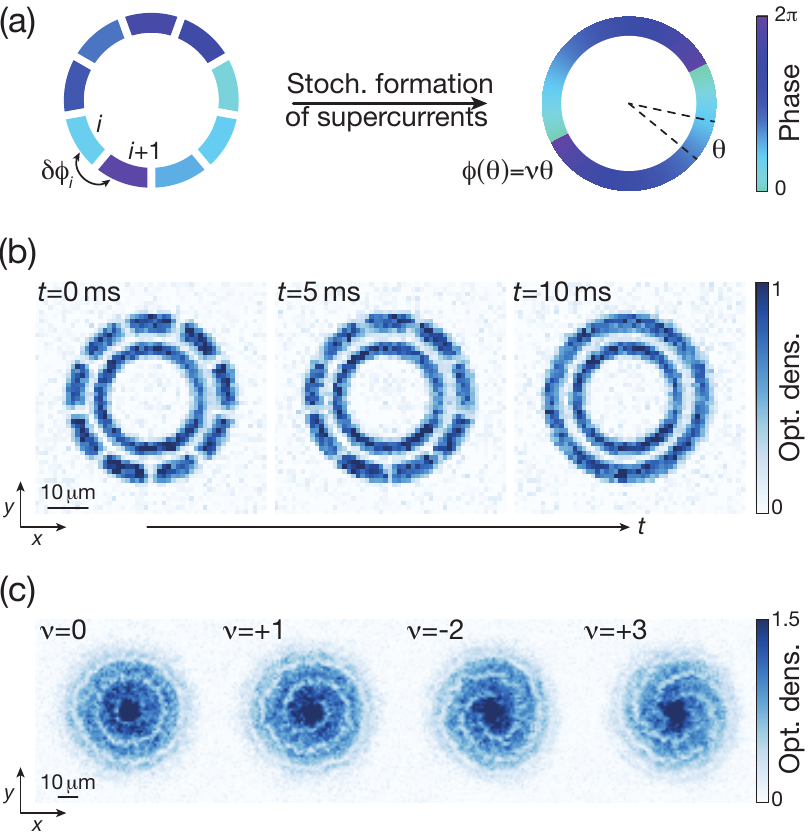}
\vspace{-0.cm} \caption{Experimental protocol. (a) Illustration of the experimental sequence. An annular trap is partitioned into $N$ segments of equal length. Uncorrelated BECs are prepared in these segments with random phase differences $\delta\phi_i$, $i=1,...,N$, between adjacent condensates. After merging into a single annular condensate, supercurrents with winding number $\nu \in \mathbb{Z}$ are formed. (b) In-situ density distribution in the ring trap for $N=9$ at different times $t$ during the merging. The outer ring has a mean radius of $19.5\,\mu$m and a width of $5\,\mu$m. The inner ring serves as a phase reference for the detection as described in the main text. It has a mean radius of $13\,\mu$m and a width of $4\,\mu$m. Each image is an average over 5 or 6 experimental realizations.
(c) Matter-wave interference after a 2D time-of-flight (TOF) of $6\,$ms. The chirality of the pattern and the number of spiral arms reveal the winding number $\nu$ of the supercurrent in the outer ring.
\label{Fig_1}}
\end{figure}

In this Letter we investigate the relaxation dynamics of up to $N=12$ uncorrelated Bose-Einstein condensates (BECs) after merging them in a ring-shaped optical trap. We measure the statistical distribution of metastable supercurrents and relate their emergence to the evolution of the phase defects generated at the boundaries of the BECs. The experimental protocol is depicted in Fig.~\ref{Fig_1}(a). Initially the condensates are characterized by random phase differences $\delta \phi_i$ ($i=1,...,N$) between condensates $i$ and $i+1$, that can lead to a net phase accumulation around the ring after merging [Fig.~\ref{Fig_1}(b)]. Due to the single-valuedness of the wavefunction, the phase winding around the ring has to be equal to $2\pi\nu$, with winding number $\nu \in \mathbb{Z}$. This correponds to the formation of supercurrents with quantized velocities, which we detect through matter-wave interference [Fig.~\ref{Fig_1}(c)] with an additional ring-shaped condensate with uniform phase \cite{Eckel:2014ff,Corman:2014cm}. Our results show that the magnitude of the supercurrent scales in quantitative agreement with the geodesic rule. This extends earlier works on the merging of two \cite{Jo:2007cd} and three \cite{Scherer:2007ee} condensates in a harmonic trap and on the dynamics of a large number of condensates in a two-dimensional (2D) optical lattice \cite{Schweikhard:2007ej}. Complementary results have been obtained with a large number of Josephson junctions, where the scaling with $N$ appears to be modified compared to the one studied in our work \cite{Carmi:2000bv}. Additionally we explore the underlying dynamics by merging pairs of neighboring condensates. First, we study it globally by monitoring the evolution of the winding-number distribution as a function of time. Secondly, we detect local phase defects and study their dynamics in a time-resolved manner. The observed relaxation timescales are compatible with the evolution of soliton-like phase defects.

The experiment started by loading a cold cloud of $1.4(2)\times10^5$ $^{87}$Rb atoms in the $|F=1, m_F=0\rangle$ state into a pancake-type dipole trap with tight harmonic confinement along the vertical direction, $\omega_z = 2 \pi \times 1.58(1)\,$kHz, and negligible confinement in the $xy$-plane \cite{Ville:2017dc,supplements}. The in-plane trap was shaped using a digital micromirror device (DMD) in direct imaging with an optical resolution of $\sim1\,\mu$m to create a uniform double-ring trap as illustrated in Fig.~\ref{Fig_1}(b). All experimental studies were performed in the outer ring, which was partitioned into several segments, while the inner ring served as a uniform phase reference for detection \cite{Eckel:2014ff,Corman:2014cm}. The distance between the segments as well as between the two rings was $2.5(2)\,\mu$m, defined as the full width at half maximum of the density dip in the measured in-situ distributions [Fig.~\ref{Fig_1}(b)]. This separation is large enough to enable the formation of uncorrelated condensates \cite{supplements}.

After $2\,$s evaporative cooling, we reached a final temperature of $T<20\,$nK, thereby entering the quasi-2D regime $k_B T < \hbar \omega_z$, with $k_B$ the Boltzmann constant and $\hbar$ the reduced Planck constant. The upper temperature limit of $20\,$nK is the smallest detectable temperature using our calibration method. This corresponds to 2D phase-space densities $\mathcal{D}=\lambda_T^2 n>80$ deeply in the superfluid regime \cite{MerminWagner}; here $n= 36(4)/\mu$m$^2$ is the 2D atomic density, $\lambda_T=\hbar \sqrt{2 \pi / (m k_B T)}$ the thermal wavelength and $m$ the mass of one atom. 

Subsequently, we merged the BECs in the outer ring within $10\,$ms by decreasing the width of the potential barriers [Fig.~\ref{Fig_1}(b)] using our dynamically configurable DMD. The velocity at which the barriers were closed was chosen small compared to the speed of sound $c_0$ in order to prevent the formation of shock waves and high-energy excitations \cite{Chang:2008ia,Meppelink:2009gr}. For our experimental parameters $c_0=\sqrt{n g_{\text{2D}}/m}\approx 1.4(1)\,$mm/s, where $g_{\text{2D}}=g_{\text{3D}}/(\sqrt{2\pi}l_z)$ is the 2D interaction parameter, $g_{\text{3D}}=4\pi\hbar^2 a/m$, $a=5.3\,$nm the scattering length and $l_z=\sqrt{\hbar/(m \omega_z)}=0.27\,\mu$m the harmonic oscillator length. 

After a typical relaxation time of $0.5\,$s, we detected the phase winding after 2D time-of-flight by releasing the in-plane confinement abruptly while keeping the vertical one. We recorded the resulting interference pattern after $6\,$ms using standard absorption imaging along the $z$-direction [Fig.~\ref{Fig_1}(c)]. The chirality of the pattern and the number of spiral arms are a direct measure of the winding number $\nu$ of the supercurrent that was formed in the outer ring \cite{Eckel:2014ff,Corman:2014cm}. In an independent calibration measurement we found that the probability of creating a supercurrent in the inner ring was $\lesssim 0.6\%$ \cite{supplements}.

\begin{figure}[t!]
\includegraphics{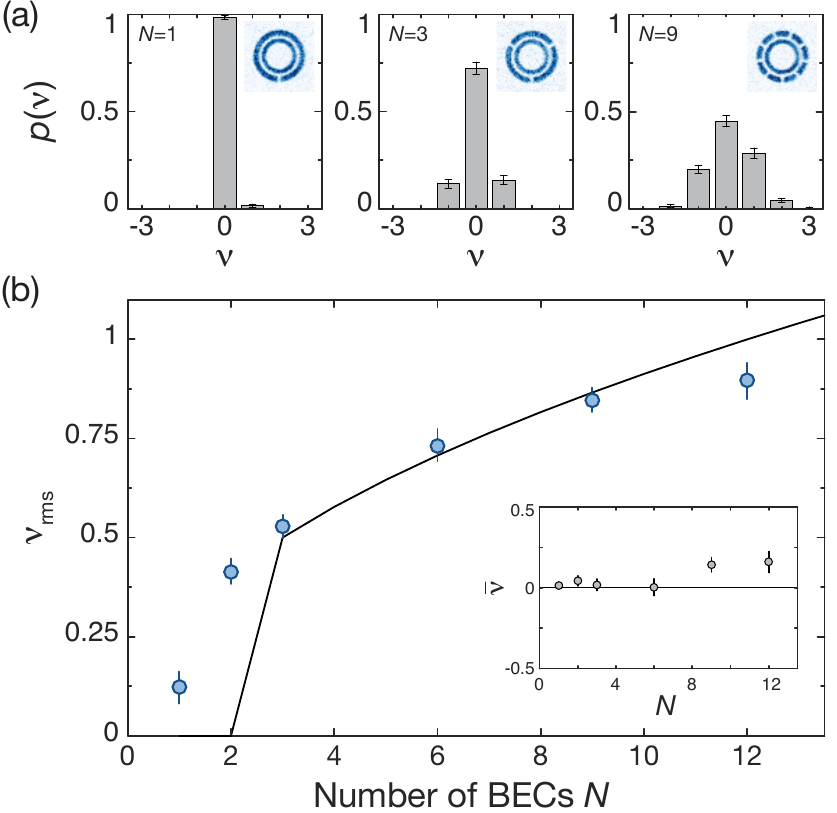}
\vspace{-0.cm} \caption{Formation of supercurrents as a function of the number of BECs $N$. (a) Probability distributions $p(\nu)$ for $N=1$, $3$ and $9$ obtained from $\mathcal{M}=202, 238$ and $388$ measurements respectively. The insets display in-situ images before the merging averaged over 4-6 realizations. (b) Measured rms-width $\nu_{\text{rms}}$ of the probability distributions as a function of $N$. Each data point consists of $\mathcal{M}>200$ independent measurements. The corresponding mean values $\bar{\nu}$ are displayed in the inset. The solid line is the predicted scaling given in Eq.~(\ref{eq:1}). All error bars display the combined uncertainty from the experimental determination of the winding number and the statistical error due to a finite number of measurements $\mathcal{M}$, which was evaluated using a bootstrapping approach.
\label{Fig_2}} 
\end{figure}

Each repetition of the experiment results in a different set of random phase differences $\delta\phi_i$ that leads to the formation of a supercurrent with winding number $\nu=\sum_{i=1}^N \delta\phi_{i}/(2\pi)$, where $-\pi < \delta\phi_i \leq \pi$. The interval for the phase differences $\delta \phi_i$ is chosen according to the geodesic rule, which expresses the fact that the system tends to minimize the absolute value of the relative phase between neighboring condensates due to energetic reasons \cite{Kibble:1995cq,Rudaz:1999kp}. By repeating the measurement $\mathcal{M}$ times we extracted the corresponding probability distributions $p(\nu)$ as illustrated in Fig.~\ref{Fig_2}(a). We observe an increase of the probability for non-zero winding numbers with $N$ resulting in a broadening of the distribution. The measured center $\bar{\nu}= \sum_\nu p(\nu) \nu$ and rms-width $\nu_{\text{rms}} = \sqrt{\mathcal{M}/(\mathcal{M}-1) \sum_{\nu} p(\nu)(\nu-\bar{\nu})^2} $ of the individual distributions are depicted in Fig.~\ref{Fig_2}(b). 

Ideally the smallest number of domains that allows for the formation of topological defects is three. In this case the probabilities $p_{\text{th}}(\nu)$ can be computed following simple arguments \cite{Bowick:1992tz,Scherer:2007ee}. There are three possible cases: if $\delta \phi_1+\delta \phi_2 > \pi$, the total sum of all phase differences has to amount to $2\pi$, if $\delta \phi_1+\delta \phi_2 < -\pi$ the total sum amounts to $-2\pi$ and for all other cases it vanishes. The resulting probabilities are $p_{\text{th}}(+1)=p_{\text{th}}(-1)=1/8$, which is compatible with our experimental results $p(+1)=0.15(2)$ and $p(-1)=0.13(2)$ displayed in Fig.~\ref{Fig_2}(a). In general the probability distribution is determined by the Euler-Frobenius distribution \cite{Janson:2013tw} and we obtain

\begin{equation}
   \nu_{\text{rms}} (N)= 
\begin{cases}
    \  0 \ ,& \text{if } N < 3    \\[7pt]
    \ \dfrac{1}{2\sqrt{3}} \sqrt{N} \ ,              & \text{if } N\geq 3 \ .
\end{cases}
\label{eq:1}
\end{equation}

\noindent The distribution is symmetric around $\nu=0$, with $\bar{\nu}=0$, which is in agreement with our experimental data obtained for small $N$ [Fig.~\ref{Fig_2}(b)]. For $N\geq9$ there seems to be a small systematic shift to positive values.

Our experimental results shown in Fig.~\ref{Fig_2}(b) are in agreement with the predicted scaling for $N \geq 3$. There is a discrepancy for $N=1$, where we measure a non-zero probability for the formation of supercurrents $p(\nu \neq 0)=1.5(8)\%$. We attribute this to phase fluctuations of the condensate due to finite temperature effects, which are enhanced for larger systems. We tested that reducing the radius of the condensate by one third significantly reduces the probability for non-zero winding numbers. For $N\geq 3$ thermal fluctuations are not expected to have a large influence because the length of the condensates is smaller. Regarding the case of $N=2$ we found that this particular configuration was very sensitive to the alignment of our trap. Small trap inhomogeneities had a significant impact on the obtained distributions.

For the largest number of condensates $N=12$ we measure slightly smaller values than expected, most likely due to an increased sensitivity to experimental imperfections and overlapping timescales. If the merging of the BECs is performed too slowly, there are two main effects that can lead to a reduction of $\nu_{\text{rms}}$. If supercurrents are already formed during the merging, their lifetime could be reduced substantially due to the presence of residual weak potential barriers \cite{Ramanathan:2011bi}. At the same time an asynchronous merging of the barriers could effectively reduce the total number of initial condensates, if the phase of neighboring condensates homogenizes before the merging is complete. We have investigated this in more detail for $N=9$ and found a significant reduction of the winding numbers for merging times larger than $50\,$ms \cite{supplements}. Both effects are expected to be more critical for increasing $N$. 

\begin{figure}[!t]
\includegraphics{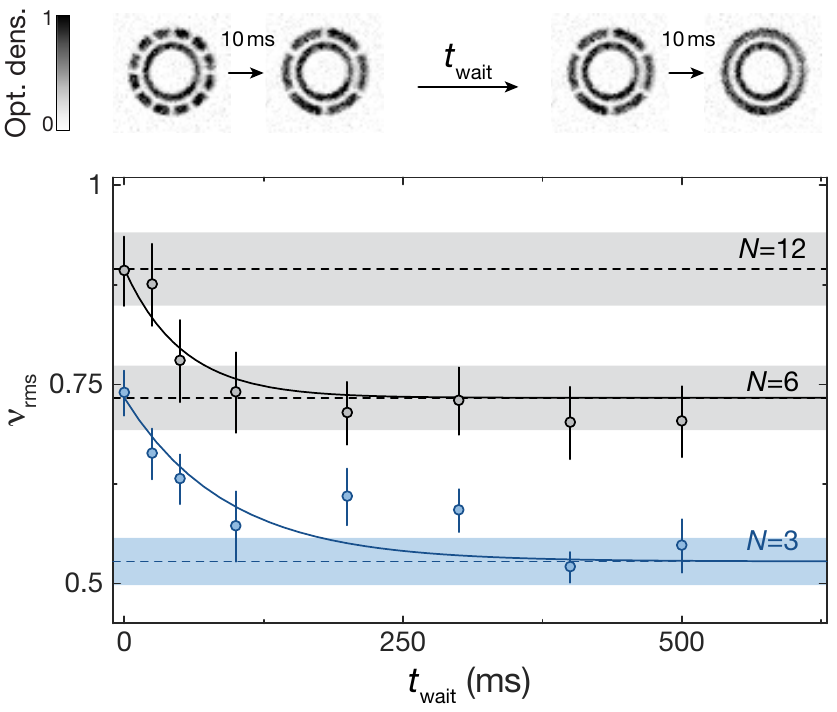}
\vspace{-0.cm} \caption{Relaxation dynamics from $N$ to $N/2$ condensates, when merging them in two successive steps. The in-situ images above the main graph illustrate the experimental sequence for $N=12$. Each image is an average over 5 individual measurements. The main graph depicts our experimental results for $N=12$ (black) and $N=6$ (blue). Each data point consists of $\mathcal{M}>200$ measurements. The corresponding mean values $\bar{\nu}$ are shown in the supplemental material. The error bars depict the uncertainty obtained from our finite number of measurements $\mathcal{M}$ and the experimental uncertainty in the determination of the winding numbers. The dashed lines indicate the measured values shown in Fig.~\ref{Fig_2}(b) and the shaded areas illustrate the corresponding error bars. The solid lines are fits of exponential functions $f_j(t_{\text{wait}})=A_j\text{e}^{-t_{\text{wait}}/\tau_j}+B_j$, $j=\{6,12\}$, to our data, where $\tau_j$ is the only free fit parameter and the other variables are determined by the dashed lines extracted from Fig.~\ref{Fig_2}(b). \label{Fig_3}} 
\end{figure} 

We typically wait $0.5\,$s after merging the condensates before detecting the supercurrents. This waiting time is short compared to the lifetime of the supercurrents in our trap \cite{supplements}. Indeed we observe no significant decay of the supercurrents for waiting times on the order of $10\,$s. On the other hand it is long enough to let the system relax to a steady state with a smoothened phase profile, without a significant number of defects in the interference pattern.

\begin{figure}[t!]
\includegraphics{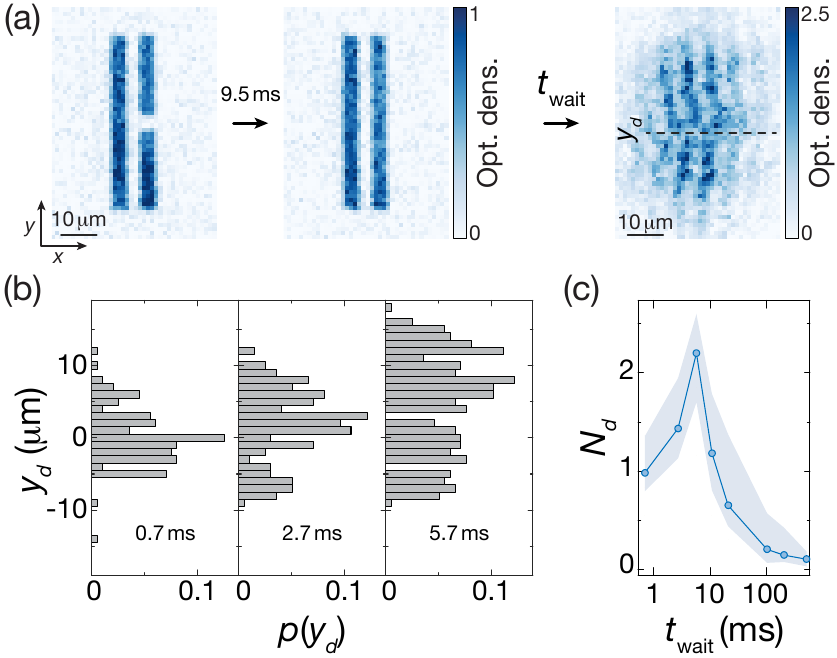}
\vspace{-0.cm} \caption{Defect dynamics. \label{Fig_4} 
(a) In-situ density distribution of two line-shaped condensates (first two images) with dimensions $50\,\mu$m $\times$ $5\,\mu$m before and after the merging (averaged over 4 individual realizations). The condensates are separated by $3\,\mu$m. After merging the condensates in $9.5\,$ms the system evolves for a variable time $t_{\text{wait}}$. Phase defects are detected by matter-wave interference after TOF (image on the right) \cite{supplements}. A typical image for $t_{\text{wait}}=0.7\,$ms is depicted on the right. The phase defect at position $y_d$ is highlighted by the dashed line.
(b) Position distribution $p(y_d)$ of the phase defects as a function of the waiting time $t_{\text{wait}}$ evaluated from 200 individual measurements. The histograms are normalized by the total number of measurements. Phase dislocations are detected, if the phase difference between neighboring pixels (corresponds to $1.16\,\mu$m in the atomic plane) is larger than 0.3$\pi$ \cite{supplements}. (c) Mean number of phase defects $N_d$ as a function of time. The data was evaluated using a threshold of 0.3$\pi$. The shaded area illustrates the sensitivity due to this analysis (upper bound: 0.16$\pi$, lower bound: 0.43$\pi$).}
\end{figure}

In order to gain a deeper insight into the underlying relaxation dynamics, we performed two separate experiments. First, we probed the evolution of the winding number distribution by merging the BECs on the ring in two successive steps. The sequence started by merging pairs of neighboring condensates within $10\,$ms to reduce the number of condensates by a factor of two, then we let the system relax for a variable time $t_{\text{wait}}$ and subsequently merged the remaining $N/2$ condensates in $10\,$ms into a single annular BEC (Fig.~\ref{Fig_3}). After an additional evolution time of $0.5\,$s we detected the probability distributions $p(\nu)$ using the detection method explained above. \\
We identify two limiting cases for the data shown in Fig.~\ref{Fig_3}. If there is no additional wait time ($t_{\text{wait}}=0$) between the two merging steps, the system has not enough time to relax and the probability distribution resembles the one discussed in Fig.~\ref{Fig_2}, where all condensates were merged in a single step. On the other hand, if $t_{\text{wait}}$ is longer than the relaxation time, the phase of neighboring condensates homogenizes after the first step, so that we effectively reduce the number of initial phase domains to $N/2$ and the distribution approaches the one for $N/2$ initial BECs merged in a single step. The measurements were performed for $N=12$ and $N=6$ and the dashed lines indicate the limiting cases explained above. In order to extract a timescale for the relaxation, we fit an exponential decay to each of the two datasets. The amplitude and offset of the fitting function are determined by the data points displayed in Fig.~\ref{Fig_2}(b). One can infer two different timescales $\tau_{12}=52(17)\,$ms and $\tau_{6}=90(30)\,$ms associated with the relaxation dynamics, which most likely depend on the spatial extent of the condensates, that differ by almost a factor of two for the two datasets. 

In a second set of measurements we focus on the microscopic relaxation dynamics via the time-resolved detection of local phase defects. We merged two condensates and probed the evolution of the phase profile through interference with a reference condensate (Fig.~\ref{Fig_4}a). The length of each condensate is comparable to the length of one segment studied in the relaxation dynamics discussed above for $N=6$. At short times ($\sim1\,$ms), we observe a phase defect in the center of the fringes, at the original position of the potential barrier (Fig.~\ref{Fig_4}b). With increasing time more phase defects appear and also start to propagate. After $5\,$ms the number of defects decays and we find an almost uniform distribution of their positions \cite{PosDistribution}. At long times ($>100\,$ms) almost all defects have disappeared in agreement with the results displayed in Fig.~\ref{Fig_3}.

We interpret the observed dynamics by the formation of dark solitons at the position of the potential barrier, whereby their shape depends on the random phase differences between neighboring condensates \cite{Burger:1999ew,Denschlag:2000cb,Becker:2008ioa}. Subsequently, the generated excitations propagate, interact with each other and eventually decay \cite{Kuznetsov:1988wr,Becker:2008ioa} to form a steady state with a smoothened phase profile (Fig.~\ref{Fig_3}b,c). Note, that the lifetime of solitonic excitations is typically short for 3D systems, but can be strongly enhanced in low-dimensional geometries \cite{Becker:2013hc,Lamporesi:2013bi,Donadello:2016gf}. The propagation speed of dark solitons depends on their depth and is at maximum equal to the speed of sound $c_0$, which is compatible with the observed relaxation timescales. The round-trip time at $c_0$ in the ring trap is about $90\,$ms for the configuration studied in Fig.~\ref{Fig_2}. 

In conclusion, we have reported the first quantitative study of the $\sqrt{N}$-scaling as predicted by the geodesic rule and show that the underlying relaxation dynamics is consistent with the formation of soliton-like defects. Future experiments could benefit from phase-imprinting techniques \cite{Burger:1999ew,Denschlag:2000cb,Becker:2008ioa} to study the dynamics in a fully deterministic manner. In particular, it would be interesting to study the dynamics as a function of temperature and geometry.

We thank W.~D.~Phillips and S.~Stringari for insightful discussions. This work was supported by DIM NanoK and ERC (Synergy UQUAM). This project has received funding from the European Union's Horizon 2020 research and innovation programme under the Marie Sklodowska-Curie grant agreement No. 703926.

%%%%%%%% Supplementary Material %%%%%%%%
\onecolumngrid
\clearpage
\begin{center}
\noindent\textbf{Supplementary Material for:}
\\\bigskip
\noindent\textbf{\large{Relaxation dynamics in the merging of \textit{N} independent condensates}} 
\\\bigskip
M.~Aidelsburger, J.~L.~Ville, R.~Saint-Jalm, S.~Nascimb\`ene, J.~Dalibard, J.~Beugnon
\\\vspace{0.1cm}
\small{\emph{Laboratoire Kastler Brossel, Coll\`ege de France, CNRS, ENS-PSL Research University, UPMC-Sorbonne Universit\'es, 11 place Marcelin-Berthelot, 75005 Paris, France}}
\end{center}
\bigskip
\bigskip
\twocolumngrid

\renewcommand{\thefigure}{S\the\numexpr\arabic{figure}-10\relax}
 \setcounter{figure}{10}
\renewcommand{\theequation}{S.\the\numexpr\arabic{equation}-10\relax}
 \setcounter{equation}{10}
 \renewcommand{\thesection}{S.\Roman{section}}
\setcounter{section}{10}

\section{Experimental sequence}
The experimental sequence started by loading a cold cloud ($>500\,$nK) of $^{87}$Rb atoms into a blue-detuned optical potential at wavelength $\lambda=532\,$nm. In the vertical direction the atoms were confined by an optical accordion with an initial lattice spacing of $11\,\mu$m. The in-plane confinement was provided by an additional laser beam that was shaped with a digital micromirror device (DMD) to engineer almost arbitrary in-plane trapping geometries. A detailed description of the experimental setup can be found in Ref.~\cite{Ville:2017dca}. 

In order to optimize the final atom number in the double-ring trap, displayed in Fig.~1(b) of the main text, we loaded the atoms into a disk-shaped trap with a radius of about $30\,\mu$m. Using the dynamical mode of our DMD we then displayed a movie that successively changed the trap from the disk to the final double-ring configuration. The movie consisted of 20 images and had a duration of $0.2\,$s. The last image of this movie abruptly introduced the potential barriers to partition the outer ring into $N$ individual segments. At the same time we compressed the optical accordion to a final lattice spacing of $5.6\,\mu$m \cite{Ville:2017dca}. This has the advantage that we can lower the power of the accordion beams, which reduces the strength of defects in the optical potential, while keeping a strong harmonic confinement along the $z$-direction.

As a next step we implemented a protocol to generate a reliable uniform phase in the inner ring, which is important for our interference-based detection technique \cite{Eckel:2014ffa,Corman:2014cma}. In order to achieve that, we introduced an additional barrier in the inner ring simultaneously with the ones in the outer ring. This prevents the formation of random supercurrents in the inner ring during evaporation \cite{Ramanathan:2011bia}. We cooled the atoms within $2\,$s by lowering the power of the dipole trap. Using an independent temperature calibration we extracted a final temperature of $T<20\,$nK. This value corresponds to the lowest detectable temperature using our calibration method.

Subsequently, we slowly removed the barrier of the inner ring by displaying a movie of 19 images with our DMD that decreased the width of the barrier gradually in $1.9\,$s from $2.5(2)\,\mu$m to zero. This procedure ensures that we have a probability for creating a supercurrent in the inner ring that is $\lesssim 0.6\%$. This value was evaluated from an independent calibration measurement explained in the following section. Subsequently we merged the segments in the outer ring within $10\,$ms by displaying a movie that consisted of 20 images on the DMD with decreasing barrier widths. We then let the system evolve for $500\,$ms and performed a 2D time-of-flight (TOF). For that we suddenly switched off the in-plane trap while keeping a slightly increased vertical confinement with frequency $\omega_z=2\pi\times2\,$kHz. After an expansion time of $6\,$ms we recorded the interference pattern along the $z$-direction using standard absorption imaging.

All in-situ images presented in this Letter were taken with partial imaging. The displayed densities correspond to $15\%$ of the total density and the imaging intensity was about $I/I_{\text{sat}}=0.2$, where $I_{\text{sat}}=1.67\,$mW$/$cm$^2$. To obtain a better contrast for the time-of-flight images we imaged the total atomic density at higher imaging intensity $I/I_{\text{sat}}=0.6$.

\begin{figure}[h!]
\includegraphics{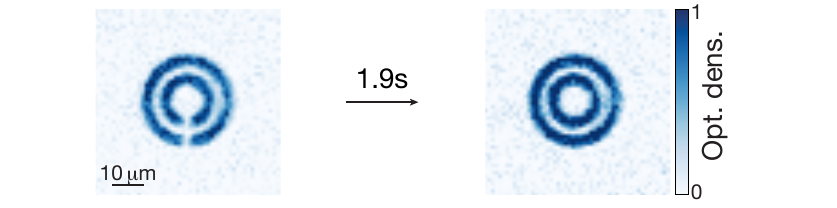}
\vspace{-0.cm} \caption{Calibration measurement for the inner ring depicted in Fig.~1(b) of the main text which serves as a uniform phase reference in our detection scheme. In-situ images of the atoms in the double-ring trap averaged over 6 and 10 individual images respectively. Initially there is a $2.5(2)\,\mu$m-wide barrier in each ring (image on the left). The image on the right shows the atomic density distribution after removing the barriers in $1.9\,$s. 
\label{Fig_S1}}
\end{figure}

\section{Phase reference of the inner ring}
To investigate the formation of supercurrents in the inner ring using the protocol described in the previous section, we loaded the atoms into a double-ring trap with slightly smaller dimensions (Fig.~\ref{Fig_S1}). The outer ring has the same dimensions as the inner ring depicted in Fig.~1(b) of the main text, with an inner radius of $11\,\mu$m and an outer one of $15\,\mu$m. The second ring is smaller with an inner radius of $5\,\mu$m and an outer one of $9\,\mu$m. The experimental sequence started by loading the atoms into this double ring potential, where we have introduced a $2.5(2)\,\mu$m-wide barrier in both rings (Fig.~\ref{Fig_S1}). Using the detection method described above, we took $\mathcal{M}=159$ measurements without observing any non-zero phase winding. In order to estimate the probability for non-zero phase windings in the ring we compute an upper bound according to the maximum likelihood estimation. According to this, the true probability $p(\nu \neq 1)$ is likely to be smaller than $ 1/159=0.6\%$ \cite{Bailey:1997a}. 

\begin{figure}[h!]
\includegraphics{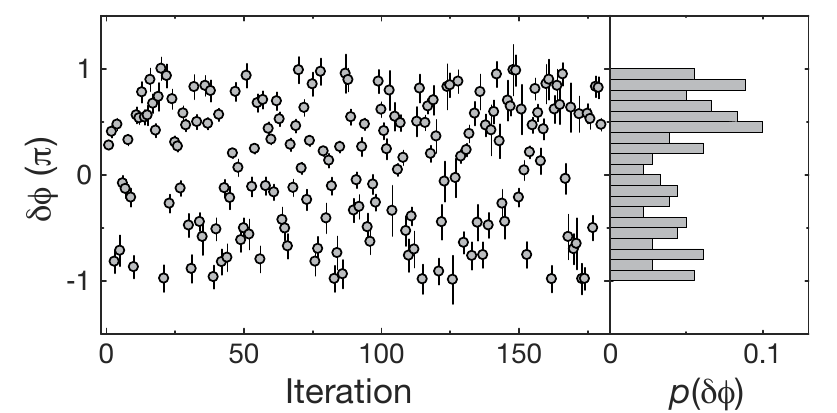}
\vspace{-0.cm} \caption{Relative phase between the two ring-shaped condensates. The relative phases $\delta\phi$ were evaluated using the data displayed in Fig.~2(a) of the main text for $N=1$ and restricting the analysis to images with $\nu=0$. The results are shown in the left panel. The corresponding histogram displaying the binned probabilities $p(\delta\phi)$ is depicted in the right panel. \label{Fig_S2}}
\end{figure}

\section{Preparation of independent BECs}
A crucial ingredient of our experimental protocol is the ability to form uncorrelated condensates in our trap via the introduction of additional potential barriers. The height of the potential is determined by the total depth of the optical potential and the width of $2.5(2)\,\mu$m is the full width at half maximum of the density dip in the in-situ distribution. In order to test the independence of the individual condensates we evaluated the relative phase $\delta\phi$ between two ring-shaped condensates with uniform phases separated by $2.5(2)\,\mu$m. For the analysis we took the data displayed in Fig.~2(a) for $N=1$ and removed all images with non-zero phase winding. For each image we computed the radial average and subtracted the background obtained by the total average of all images. The resulting density modulation was fitted with a sinusoidal function $f(r)=A\,\text{sin}(kr+\delta\phi)$ to determine the phase of the interference pattern $\delta\phi$ (Fig.~\ref{Fig_S2}). The period $k$ was set to the mean period obtained from the individual fit results and the amplitude $A$ and the phase offset $\delta\phi$ were free fit parameters. In total we have evaluated $\mathcal{M}=180$ images, where we have excluded those, where the fit error of $\delta\phi$ exceeded $0.25\pi$. The obtained distribution (Fig.~\ref{Fig_S2}) is characterized by the mean values: $\langle \textrm{cos}(\delta \varphi)\rangle=0.11(5)$ and $\langle \textrm{sin}(\delta \varphi)\rangle=-0.16(5)$. Ideally these values would vanish for a perfectly random distribution. We attribute the residual deviations to a systematic error in our fitting protocol, which is mainly caused by the non-uniform shape of the overall time-of-flight distribution. The assumption of initially uncorrelated BECs is further supported by additional experimental tests, where we have measured the probability distribution for $N=9$ for different thicknesses of the potential barriers and did not observe any significant change when varying the width by $\sim \pm 20\,\%$.

\begin{figure}[h!]
\includegraphics{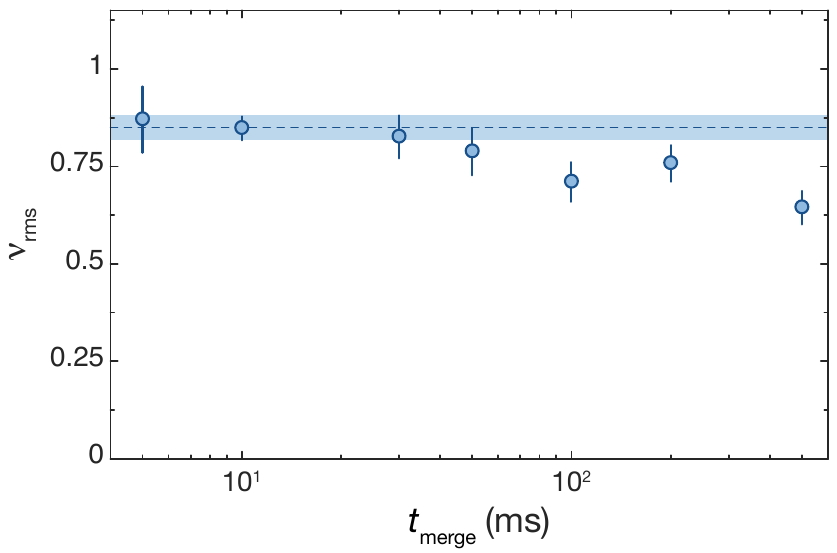}
\vspace{-0.cm} \caption{Width of the distribution for $N=9$ as a function of the merging time $t_{\text{merge}}$. Each data point was obtained from a set of 50-100 individual measurements. The data point at $10\,$ms corresponds to the merging time we used for the data presented in the main text and consists of $\mathcal{M}=388$ measurements. The vertical error bars denote the statistical uncertainty due to a finite number of measurements $\mathcal{M}$ and the uncertainty in the determination of the winding numbers. \label{Fig_S3}}
\end{figure}

\section{Width of the distribution for different merging times}
For the measurements presented in the main text we merged the condensates within $10\,$ms. This time was chosen between two limiting regimes. If the condensates are merged too fast, we may create shock waves and high-energy excitations \cite{Chang:2008iaa,Meppelink:2009gra} as opposed to the quasi-adiabatic generation of supercurrents we want to probe in the experiment. On the other hand if the merging is performed too slowly, the formation of supercurrents with non-zero phase winding might be reduced due to an asynchronous closing of the individual barriers since the phase between neighboring condensates may homogenize before the merging is complete. Moreover, the lifetime of the supercurrents may be reduced in the presence of weak potential barriers around the ring when the supercurrents are formed during the merging \cite{Ramanathan:2011bia}. Experimentally we investigated these effects for $N=9$ initial condensates. We found that varying the merging time by about one order of magnitude between $5\,$ms and $50\,$ms did not significantly influence our experimental results (Fig.~\ref{Fig_S3}). When the barriers were removed abruptly instead we observed a significant increase in the width of the distribution $\nu_{\text{rms}}$ and the interference pattern exhibited many phase defects.

\section{Lifetime of supercurrents}
\begin{figure}[h!]
\includegraphics{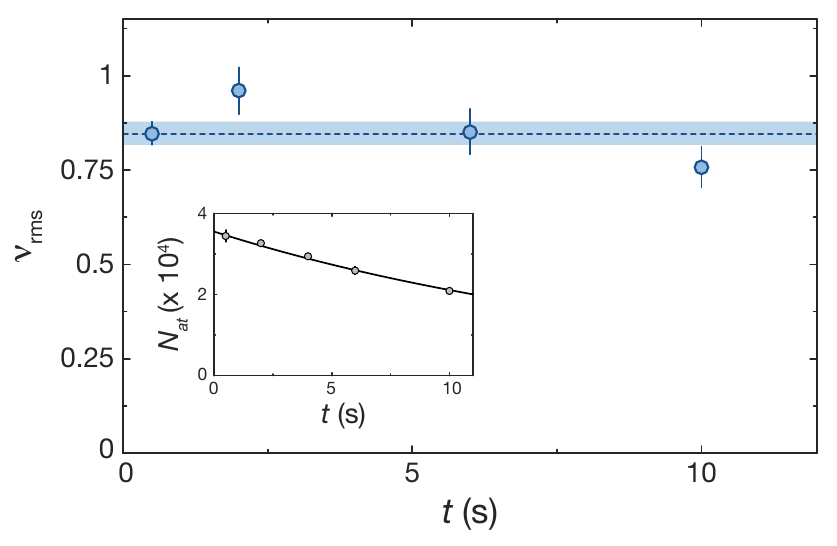}
\vspace{-0.cm} \caption{Lifetime of supercurrents and atoms for $N=9$ segments. The data points show the measured rms-width $\nu_{\text{rms}}$ as a function of time $t$ after merging the condensates. Each data point was evaluated from $\mathcal{M}>100$ images. The data point at $t=0.5\,$s corresponds to the typical experimental sequence with $\mathcal{M}=388$ [Fig.~2(a) in the main text]. It is further highlighted by the dashed line, the shaded area represents its uncertainty. All error bars display the combined error of statistical uncertainties and uncertainties in the analysis of the interference patterns. The inset displays the corresponding atom numbers. For each data point we evaluated between 2 and 13 measurements and the vertical error bars depict the standard deviation. The solid line is the fit of an exponential function $f(t)=A\text{e}^{-t/\tau}$ to our data, resulting in $\tau=19(1)\,$s.\label{Fig_S4}}
\end{figure}

For the measurements reported in the main text we typically wait $0.5\,$s after merging the condensates. We have studied the lifetime of the supercurrents using the same experimental sequence and increasing the waiting time before detection. The results are displayed in Fig.~\ref{Fig_S4}. We observed no significant decay of the supercurrents for waiting times that are on the order of the atomic lifetime (inset Fig.~\ref{Fig_S4}). This timescale is large compared to timescale of a typical experimental sequence.

\section{Center values for the relaxation measurements}

\begin{figure}[h!]
\includegraphics{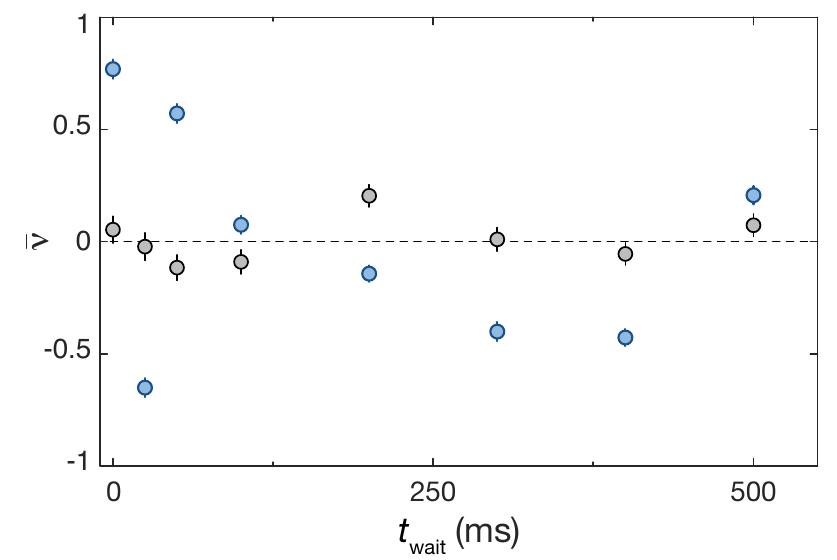}
\vspace{-0.cm} \caption{Center values $\bar{\nu}$ of the distributions corresponding to the widths $\nu_{\text{rms}}$ shown in Fig.~3 of the main text for $N=12$ (black) and $N=6$ (blue) respectively. Each data point consists of $\mathcal{M}>200$ measurements. The vertical error bars depict the standard deviation obtained from statistical uncertainties and the experimental uncertainty in the determination of the winding numbers.\label{Fig_S5}}
\end{figure}

In Fig.~\ref{Fig_S5} we display the mean values $\bar{\nu}$ of the distributions corresponding to the data displayed in Fig.~3 of the main text, where we have studied the relaxation from $N$ to $N/2$ condensates, when merging them in two successive steps. For $N=6$ initial segments we observe consistently larger asymmetries in the distributions and they seemed to be reproducible over the course of several days. In the case of $N=12$ the observed asymmetries are consistent with statistical fluctuations.

\section{Experimental sequence and data analysis for the merging of two condensates}

The experimental sequence for the study of the dynamics of the defects was as follows: We loaded a rectangle of size $50 \times 30\,\mu$m with thermal atoms and dynamically reduced its width to $13\,\mu$m within $0.2\,$s. A separation of width $3\,\mu$m was abruptly introduced in the middle of the rectangle to create two lines of dimensions $50 \times 5\,\mu$m. One of the lines was further cut into two lines of length $23.5\,\mu$m. The other line serves as a phase reference for the detection. The system was then cooled down within $2\,$s to the same temperature as given in the main text. After an equilibration time of $1\,$s the separation between the two $23\,\mu$m-long lines was removed within $9.5\,$ms, following the same procedure as discussed in the previous sections. We then let the system relax for a variable time $t_{\text{wait}}$ and subsequently detected the interference pattern by performing a 2D-TOF of $3\,$ms (as introduced in the section \textit{Experimental sequence}) followed by a 3D-TOF of $3\,$ms during which we removed all confining potentials.

\begin{figure*}
\includegraphics{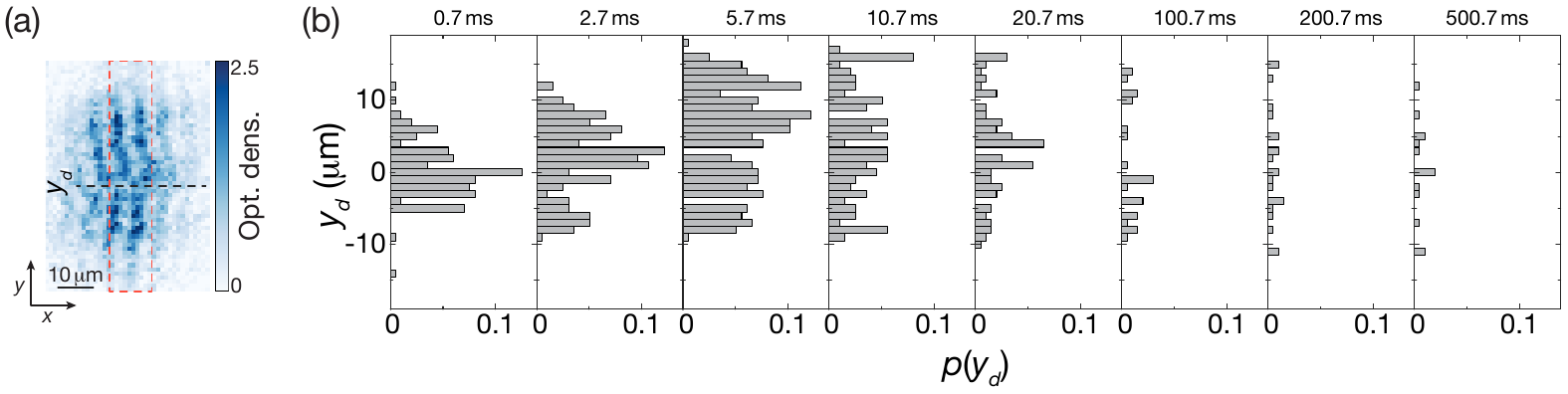}
\vspace{-0.cm} \caption{(a) Interference pattern after an evolution time $t_{\text{wait}}=0.7\,$ms detected after a 2D-TOF of $3\,$ms plus a 3D-TOF of $3\,$ms. The red-dashed lines mark the region-of-interest for the data analysis. The phase dislocation at $x_d$ his highlighted by a black dashed line. (b) Position distributions $p(y_d)$ of the phase dislocations for various evolution times $t_{\text{wait}}$. The histograms are normalized by the total number of images. For each time $t_{\text{wait}}$ we have taken about 200 individual measurements. For this data the threshold for the analysis described in the text is $\varphi_{\text{c}}=0.3\pi$. \label{Fig_S6}}
\end{figure*}

Phase defects in the interference pattern were analyzed in the following way: We chose a region-of-interest in the center of the cloud in the $x$-direction, which represents two periods of the fringes (Fig.~\ref{Fig_S6}a) and performed a sliding average of the image along $y$ to smoothen the profile. The wavevector of the fringes along $x$ was determined via Fourier transform. For each position in the $y$-direction, the modulus of this Fourier coefficient gives us the visibility of the fringes and its phase is the local relative phase of the two lines at this position. The dislocations can be found at the positions where both the visibility drops and the phase jumps. More precisely we were looking for coincidences of minima of the visibility and maxima of the absolute value of the phase gradient along the $y$-direction. We removed an overall smooth gradient of the phase across the whole cloud to avoid any systematic bias between positive and negative phase jumps. We evaluated the data for different threshold values $\varphi_{\text{c}}$ between neighboring pixels (this corresponds to an effective pixel size of $1.16\,\mu$m in the atomic plane), where a phase dislocation is detected above this threshold. All phase dislocations below this value are discarded. This threshold is needed in order to avoid the detection of false dislocations. The extracted positions $y_d$ of the detected phase dislocations are shown in the histograms (Fig.~\ref{Fig_S6}b), which are normalized by the total number of individual measurements.

In order to evaluate the total number of defects $N_d$ (Fig.~4c in the main text) as a function of time we assume that phase jumps occur with an equal probability in the interval $(-\pi,\pi]$. The threshold value $\varphi_{\text{c}}$ will artificially reduce the number of detected defects by a factor $(1-\varphi_{\text{c}}/\pi)$. Therefore we use this as a correction factor and rescale our data accordingly. This is certainly true for short times but it is not necessarily the case for long times because the relaxation dynamics most likely depend on the initial phase difference between the two condensates. Nonetheless it helps us to estimate the error we make in the data analysis due to the finite threshold value $\varphi_{\text{c}}$. In Fig.~4c of the main text we display the results for different threshold values between $0.16\pi$ and $0.43\pi$.


\begin{thebibliography}{37}
\bibitem{MerminWagner}
According to the Mermin-Wagner theorem there is no true long-range order in low-dimensional systems, however, for our experimental parameters the characteristic correlation length exceeds the size of the condensate by several orders of magnitude, which enables the formation of a Bose-Einstein condensate.
\bibitem{PosDistribution}
The asymmetry in the position distribution of the phase defects (third panel of Fig.~\ref{Fig_4}b) is due to a bias in the contrast of the interference fringes.
\end{thebibliography}
\end{document}